\documentclass[12pt]{article}

\usepackage{graphicx}

\begin{document}

\title{Exact solution of the Schr\"odinger equation with a Lennard-Jones potential}



\author{J. Sesma\thanks{e-mail: javier@unizar.es}\\  \   \\
Departamento de F\'{\i}sica Te\'{o}rica, \\ Facultad de Ciencias,
\\ 50009 Zaragoza, Spain. \\  \ }

\date{ }

\maketitle

\begin{abstract}
The Schr\"odinger equation with a Lennard-Jones potential is solved by using a procedure that treats in a rigorous way the irregular singularities at the origin and at infinity. Global solutions are obtained thanks to the computation of the connection factors between Floquet and Thom\'e solutions. The energies of the bound states result as zeros of a function defined by a convergent series whose successive terms are calculated by means of recurrence relations. The procedure gives also the wave functions expressed either as a linear combination of two Laurent expansions, at moderate distances, or as an asymptotic expansion, near the singular points. A table of the critical intensities of the potential, for which a new bound state (of zero energy) appears, is also given.
\end{abstract}

\section{Introduction}

The interaction between two atoms is frequently represented by means of a Lennard-Jones potential,
\begin{equation}
V(r) = \frac{\hbar^2}{2\,m\,r_{\rm e}^2}\,\lambda\left[\left(\frac{r_{\rm e}}{r}\right)^{12}-2\left(\frac{r_{\rm e}}{r}\right)^{6}\right], \label{i1}
\end{equation}
alone or with addition of some corrections. In this expression $m$ is the reduced mass of the system of two atoms, $r_{\rm e}$ is the equilibrium distance (minimum of $V(r)$) and $\lambda$ is a dimensionless parameter accounting for the intensity of the interaction. Both $r_{\rm e}$ and $\lambda$ are empirically adjusted for each particular kind of interacting atoms. Other classical interatomic potentials, like the Morse, Rydberg or Buckingham ones, can be simulated, as shown by Lim \cite{lim}, by one of the Lennard-Jones type.

Given a diatomic system and assumed a certain potential to represent the interaction, one is interested, from a theoretical point of view, mainly on the determination of its spectrum of energies, to be compared with the experimentally observed bound states. Nevertheless, in many cases, one needs to know also the corresponding wave functions in order to compute the expected values of quantities that may be obtained in the experiment. A large variety of algebraic methods are discussed in the monographs by Fern\'andez and Castro \cite{fer1} and by Fern\'andez \cite{fer2}. References to later developments can be found in recently published papers \cite{oye1,acka,hamz,oye2}.
Numerical methods have been developed, among others, by Simos and collaborators \cite{vig1,vig2,vig3,vig4}. An extensive bibliography concerning those methods can be found in Section 2 of a recent paper \cite{simo}. Except for a few familiar potentials, for which the differential equation can be solved exactly \cite{flug}, those methods provide only with approximate values of the energies and wave functions. This may be sufficient in most of cases. However, due to the strong singularity at the origin of the Schr\"odinger equation with a Lennard-Jones potential, those approximate methods cannot represent faithfully the behaviour of the wave function in the neighbourhood of the origin. This fact, besides of being unsatisfactory from a mathematical point of view, may constitute a serious inconvenient for the computation of the expected values of certain operators.

The purpose of this paper is to call the attention of users of the Lennard-Jones potential towards a method of solution of the Schr\"odinger equation that is able to give the correct behaviour of the wave function in the neighbourhood of the origin and the infinity, the two singular points of the differential equation. The method is exact, free of approximations, although errors due to the computational procedure are unavoidable. But these errors can be reduced by increasing the number of digits carried along the calculations.

We present, in the next Section, fundamental sets of solutions of the Schr\"odinger equation that serve as a basis to express the physical solution. The requirement of a regular behaviour of this solution at the singular points establishes a condition, in terms of the connection factors, to be fulfilled by the energies of the bound states. The procedure to determine the connection factors is explained in Section 3. The energies of the bound states in a potential of intensity $0\leq\lambda\leq 100$ are shown in Figure 1. Expressions of the corresponding wave functions are given in Section 4. As $\lambda$ increases, new bound states appear. We denote as {\it critical} those values of $\lambda$ for which a state of zero energy exists. In Section 5, a method is suggested to find those critical intensities, which are reported in Table 5. Section 6 contains some pertinent comments. Finally, we recall, in an Appendix, a procedure to solve the nontrivial problem of finding the Floquet solutions.

\section{Solutions of the Schr\"odinger equation}

For a given energy $E$ and angular momentum $l$, the Schr\"odinger equation for the reduced radial wave function, $R(r)$, of a particle of mass $m$ in the potential $V(r)$, given in Eq. (\ref{i1}), reads
\begin{equation}
-\frac{\hbar^2}{2m}\left(\frac{d^2R(r)}{dr^2}-\frac{l(l+1)}{r^2}R(r)\right)+V(r)\,R(r)=E\,R(r).  \label{ii1}
\end{equation}
As usual, we will express the solutions of this differential equation in terms of dimensionless radial variable, $z$,  and energy parameter, $\varepsilon$, defined by
\begin{equation}
z\equiv\frac{r}{r_{\rm e}}, \qquad \varepsilon\equiv\frac{2mr_{\rm e}^2}{\hbar^2}E. \label{ii2}
\end{equation}
For the radial wave function in terms of the new variable we will use
\begin{equation}
w(z)\equiv R(r).  \label{ii3}
\end{equation}
Then, the Schr\"odinger equation becomes
\begin{equation}
-z^2\,\frac{d^2w(z)}{dz^2}+\left(\lambda\,z^{-10}-2\lambda\,z^{-4}+l(l+1)-\varepsilon\,z^2\right)\,w(z) = 0. \label{ii4}
\end{equation}
This differential equation presents two irregular singular points: one of rank 5 at the origin, an another of rank 1 at infinity. The physical solution must be regular at both singular points. To express this solution, we find convenient to consider three different fundamental systems of solutions.

\subsection{Floquet solutions}

Except for certain particular values of the parameters $\lambda$ and $\varepsilon$, that we exclude from this discussion, there are  two independent {\em Floquet} or {\em multiplicative} solutions expressed as Laurent power series of the form
\begin{equation}
w_i=z^{\nu_i}\sum_{n=-\infty}^{\infty} c_{n,i}\,z^n, \quad \mbox{being} \; \sum_{n=-\infty}^{\infty} |c_{n,i}|^2<\infty, \quad i=1,2. \label{ii5}
\end{equation}
The indices $\nu_i$ are not uniquely defined. They admit addition of any integer (with an adequate relabeling of the coefficients).
In the general case, the indices $\nu_i$ and the coefficients $c_{n,i}$ may be complex. The requirement that $w_i(z)$ be a solution of (\ref{ii4}) gives the recurrence relation
\begin{equation}
\varepsilon\,c_{n-2,i}+\left[(n\! +\! \nu_i)(n\! -\! 1\! +\! \nu_i)-l(l\! +\! 1)\right]\,c_{n,i} +2\lambda\,c_{n+4,i}-\lambda\,c_{n+10,i}=0\,,
 \label{ii6}
\end{equation}
The solution of this difference equation is not trivial. It can be treated as a nonlinear eigenvalue problem. In Appendix A we show an implementation of the Newton method to determine the indices $\nu_i$ and the coefficients $c_{n,i}$.

\subsection{Thom\'e solutions for large values of $z$}

There are two other independent solutions characterized by their behaviour for $z\to\infty$, namely
\begin{equation}
w_j(z)\sim\exp\left(\alpha_{j}\,z\right) \,\sum_{m=0}^\infty a_{m,j}\,z^{-m}, \qquad a_{0,j}\neq 0, \qquad j=3, 4. \label{ii7}
\end{equation}
It can be easily checked, by taking
\begin{equation}
\alpha_j=\sqrt{-\varepsilon}   \label{ii8}
\end{equation}
and coefficients $a_{m,j}$ given by (omitting the second subindex, $j$)
\begin{equation}
a_{0}=1,\quad 2\alpha\,m\,a_{m}=\left[m(m-1)-l(l+1)\right]a_{m-1}+2\lambda\,a_{m-5}-\lambda\,a_{m-11}\,,  \label{ii9}
\end{equation}
that the right hand side of Eq. (\ref{ii7}) is a solution of the differential equation (\ref{ii4}). In fact, it is a formal solution, as the series is an asymptotic one that does not converge in general. The two values of the subindex $j$ in Eq. (\ref{ii7}) correspond to the two possible values of the right hand side of Eq. (\ref{ii8}). In the case of negative energies, we adopt the convention
\begin{equation}
\alpha_3=-\sqrt{-\varepsilon},\qquad  \alpha_4=+\sqrt{-\varepsilon}.  \label{ii10}
\end{equation}
Accordingly, $w_3(z)$ is physically acceptable, as it vanishes at infinity, whereas $w_4(z)$ diverges and, therefore, should be eliminated from the physical solution. In the case (not to be considered in this paper) of positive energies, both $w_3(z)$ and $w_4(z)$ are oscillating solutions and correspond to incoming and outgoing waves.

\subsection{Thom\'e solutions near the origin}

In the neighbourhood of the origin, the role analogous to that of $w_3$ and $w_4$ at infinity is played by two other solutions, $w_5$ and $w_6$, such that, for $z\to 0$,
\begin{equation}
w_k(z)\sim\exp\left(\beta_{k}\,z^{-5}/5\right)\,z^{\rho_k}\,\sum_{m=0}^\infty b_{m,k}\,z^{m}, \qquad b_{0,k}\neq 0, \qquad k=5, 6. \label{ii11}
\end{equation}
Substitution of these expressions in Eq. (\ref{ii4}) gives for the coefficients in the exponents
\begin{equation}
\beta_k=\sqrt{\lambda}, \qquad  \rho_k=3, \label{ii12}
\end{equation}
and for the coefficients in the series (omitting the second subindex, $k$)
\begin{equation}
2\beta\,m\,b_m=2\lambda\,b_{m-1}+\left[(m-3)(m-2)-l(l+1)\right]\,b_{m-5}+\varepsilon\,b_{m-7},  \label{ii13}
\end{equation}
a recurrence relation that allows one to obtain the $b_{n,k}$ by starting with
\begin{equation}
b_{0,k}=1.  \label{ii14}
\end{equation}
The two solutions correspond to the two possible values of the right hand side of the first of Eqs. (\ref{ii12}). By convention we take
\begin{equation}
\beta_5=-\sqrt{\lambda}, \qquad  \beta_6=+\sqrt{\lambda}. \label{ii15}
\end{equation}
Then, $w_5$ is acceptable, from the physical point of view, whereas $w_6$ should be discarded.

\subsection{The physical solution}

As the solutions $w_1$ and $w_2$ of the differential equation constitute a fundamental system, any solution can be written as a linear combination of them. In particular, the physical solution would be
\begin{equation}
w_{\rm phys}(z)=A_1\,w_1(z)+A_2\,w_2(z),  \label{ii16}
\end{equation}
with constants $A_1$ and $A_2$, to be determined, such that $w_{\rm phys}(z)$ becomes regular at the origin and at infinity. To impose this condition we need to know the behaviour of $w_1$ and $w_2$ at the singular points. In other words, we need to calculate the {\em connection factors} $T$ defined by
\begin{eqnarray}
w_i(z) &\sim& T_{i,3}\,w_3(z)+T_{i,4}\,w_4(z), \qquad  \mbox{for}\quad z\to\infty, \qquad i=1, 2, \label{ii17}  \\
w_i(z) &\sim& T_{i,5}\,w_5(z)+T_{i,6}\,w_6(z), \qquad  \mbox{for}\quad z\to 0, \qquad i=1, 2.  \label{ii18}
\end{eqnarray}
In terms of them, the behaviour of the physical solution in the neighbourhood of the singular points would be
\begin{eqnarray}
w_{\rm phys}(z) &\sim& (A_1\,T_{1,3}+A_2\,T_{2,3})\,w_3(z)+(A_1\,T_{1,4}+A_2\,T_{2,4})\,w_4(z)\,, \quad  \mbox{for}\quad z\to\infty,  \nonumber  \\
w_{\rm phys}(z) &\sim& (A_1\,T_{1,5}+A_2\,T_{2,5})\,w_5(z)+(A_1\,T_{1,6}+A_2\,T_{2,6})\,w_6(z)\,, \quad  \mbox{for}\quad z\to 0.   \nonumber
\end{eqnarray}
The regularity of the physical solution at the singular points is guaranteed if $A_1$ and $A_2$ are chosen in such a way that
\begin{equation}
 A_1\,T_{1,4}+A_2\,T_{2,4}=0 \qquad \mbox{and} \qquad A_1\,T_{1,6}+A_2\,T_{2,6}=0\,,  \label{ii19}
\end{equation}
which is possible if and only if
\begin{equation}
T_{1,4}\,T_{2,6}-T_{2,4}\,T_{1,6}=0\,.  \label{ii20}
\end{equation}
For given values of the parameters of the potential, the left hand side of this equation is a function of $\varepsilon$ whose zeros correspond to the values of the energies of the bound states. Equation (\ref{ii20}) is, therefore, the quantization condition. Solving it requires to know the connection factors.
We present in the next Section our procedure to determine them.

\section{The connection factors}

Let us design by $\mathcal{W}[f,g]$ the Wronskian of two functions $f$ and $g$,
\begin{equation}
\mathcal{W}[f,g](z)=f(z)\,\frac{dg(z)}{dz}-\frac{df(z)}{dz}\,g(z)\,.   \label{iii1}
\end{equation}
Then, from Eqs. (\ref{ii17}) and (\ref{ii18}), one obtains immediately
\begin{eqnarray}
T_{i,3}&=&\frac{\mathcal{W}[w_i,w_4]}{\mathcal{W}[w_3,w_4]}, \qquad T_{i,4}=\frac{\mathcal{W}[w_i,w_3]}{\mathcal{W}[w_4,w_3]},\qquad i=1, 2, \label{iii2}  \\
T_{i,5}&=&\frac{\mathcal{W}[w_i,w_6]}{\mathcal{W}[w_5,w_6]}, \qquad T_{i,6}=\frac{\mathcal{W}[w_i,w_5]}{\mathcal{W}[w_6,w_5]},\qquad i=1, 2.  \label{iii3}
\end{eqnarray}
All Wronskians in these equations are independent of $z$. Those in the denominators can be calculated directly to obtain
\begin{eqnarray}
\mathcal{W}[w_3,w_4]&=&-\mathcal{W}[w_4,w_3]=2\alpha_4\,a_{0,3}\,a_{0,4}=2\sqrt{-\varepsilon}, \label{iii4}  \\
\mathcal{W}[w_5,w_6]&=&-\mathcal{W}[w_6,w_5]=-2\beta_6\,b_{0,5}\,b_{0,6}=-2\sqrt{\lambda}.  \label{iii5}
\end{eqnarray}
The calculation of the numerators is not so easy. In a former paper \cite{gom1} we suggested a procedure that has been used to find the bound states in a spiked harmonic oscillator \cite{gom2}. For convenience of the reader, we recall here the procedure, adapted to the present problem.

We consider firstly the Wronskians of each one of the Floquet solutions with the two Tom\'{e} solutions at infinity, $\mathcal{W}[w_i,w_j]$ ($i=1, 2,\; j=3, 4$). Let us introduce the auxiliary functions
\begin{equation}
u_{i,j}=\exp\left(-\alpha_j\,z/2\right)\,w_i, \qquad u_j=\exp\left(-\alpha_j\,z/2\right)\,w_j\,.  \label{iii6}
\end{equation}
Obviously,
\begin{equation}
\mathcal{W}[u_{i,j},u_j]= \exp\left(-\alpha_j\,z\right)\,\mathcal{W}[w_{i},w_j]\,.  \label{iii7}
\end{equation}
Both sides of this equation obey the first order differential equation
\begin{equation}
y^\prime=-\alpha_j\,y\,.  \label{iii8}
\end{equation}
A direct computation of the left hand side of Eq. (\ref{iii7}), by using the definitions (\ref{iii6}) and the expansions (\ref{ii5}) and (\ref{ii7}), gives the doubly infinite series
\begin{equation}
\mathcal{W}[u_{i,j},u_j]\sim \sum_{n=-\infty}^\infty \gamma_n^{(i,j)}\,z^{n+\nu_i}\,,  \label{iii9}
\end{equation}
whose coefficients
\begin{equation}
\gamma_n^{(i,j)}=\sum_{m=0}^\infty a_{m,j}\big(\alpha_j\,c_{n+m,i}-(n+2m+1+\nu_i)\,c_{n+m+1,i}\big)  \label{iii10}
\end{equation}
are solution of the first order difference equation
\begin{equation}
(n+1+\nu_i)\,\gamma_{n+1}^{(i,j)}+\alpha_j\,\gamma_n^{(i,j)} = 0\,.  \label{iii11}
\end{equation}
An expansion of the right hand side of Eq. (\ref{iii7}), analogous to that in (\ref{iii9}), can be obtained by making use of the so-called Heaviside's exponential series \cite{hard}
\begin{equation}
\exp (t)\sim \sum_{-\infty}^\infty \frac{t^{n+\delta}}{\Gamma (n+1+\delta)}\,, \qquad |\arg (t)|<\pi, \qquad \delta\; \mbox{arbitrary}.  \label{iii12}
\end{equation}
By taking $t=-\alpha_j\,z$ and choosing $\delta=\nu_i$, one gets an expansion,
\begin{equation}
\exp (-\alpha_j\,z)\sim \sum_{-\infty}^\infty \frac{(-\alpha_j)^{n+\nu_i}}{\Gamma (n+1+\nu_i)}\,z^{n+\nu_i}\,,   \label{iii13}
\end{equation}
in series of the same powers of $z$ as in (\ref{iii9}) with coefficients obeying the same first order difference equation,
\begin{equation}
(n+1+\nu_i)\,\frac{(-\alpha_j)^{n+1+\nu_i}}{\Gamma (n+2+\nu_i)}+\alpha_j\,\frac{(-\alpha_j)^{n+\nu_i}}{\Gamma (n+1+\nu_i)} = 0\,.  \label{iii14}
\end{equation}
Both solutions $\{\gamma_n^{(i,j)}\}$ and $\{(-\alpha_j)^{n+\nu_i}/\Gamma (n+1+\nu_i)\}$ of the difference equation must be related by a multiplicative constant that, in view of Eq. (\ref{iii7}), shold be $\mathcal{W}[w_{i},w_j]$. Therefore,
\begin{equation}
\mathcal{W}[w_{i},w_j]=\frac{\Gamma (n+1+\nu_i)}{(-\alpha_j)^{n+\nu_i}}\,\gamma_n^{(i,j)}\,,   \label{iii15}
\end{equation}
an expression that, together with  Eq. (\ref{iii4}), would allow one to calculate the connection factors given by Eq. (\ref{iii2}). Nevertheless, the validity of Eq. (\ref{iii15}) is subordinate to the fulfilment of the condition $|\arg (-\alpha_j\,z)|<\pi$, necessary for the validity of Eq. (\ref{iii13}). Such condition is satisfied in the case $j=3$, as, for $z\in[0,+\infty)$, $\arg(-\alpha_3\,z)=0$. There is no difficulty in computing $T_{i,4}$ by substituting, in  the second of Eqs. (\ref{iii2}),
\begin{equation}
\mathcal{W}[w_{i},w_3]=\frac{\Gamma (n+1+\nu_i)}{(-\alpha_3)^{n+\nu_i}}\,\gamma_n^{(i,3)}\,.   \label{iii16}
\end{equation}
In the case $j=4$, instead, the above mentioned condition is not satisfied and Eq. (\ref{iii15}) is not valid for $z\in[0,+\infty)$. In fact, the positive real semiaxis is a Stokes ray for $T_{i,3}$, that should be taken as the average
\begin{equation}
T_{i,3}=\frac{1}{2}(T_{i,3}^++T_{i,3}^-)  \label{iii17}
\end{equation}
of its values in the sectors separated by the ray. Equivalently, one may define
\begin{equation}
\mathcal{W}[w_{i},w_4]=\frac{1}{2}\big(\mathcal{W}[w_{i},w_4]^+ + \mathcal{W}[w_{i},w_4]^-\big)\,,   \label{iii18}
\end{equation}
an average of the Wronskians for $z$ slightly above and below the positive real semiaxis. The result is
\begin{equation}
\mathcal{W}[w_{i},w_4]=(-1)^n\,\cos (\nu_i\pi)\, \frac{\Gamma (n+1+\nu_i)}{(\alpha_4)^{n+\nu_i}}\,\gamma_n^{(i,4)}\,.  \qquad i=1, 2  \label{iii19}
\end{equation}
This equation  provides with the needed value of the numerator in the first of Eqs. (\ref{iii2}).

The procedure to calculate the Wronskians, $\mathcal{W}[w_i,w_k]$, ($i=1, 2, k=5, 6$) of each one of the Floquet solutions with the two Thom\'e solutions at the origin is analogous to that just described, with the unavoidable differences due to the fact that the singularity at the origin is of rank five, whereas it was of rank one at infinity. The auxiliary functions are now
\begin{equation}
v_{i,k}=\exp\left(-\beta_k\,z^{-5}/10\right)\,w_i, \qquad v_k=\exp\left(-\beta_k\,z^{-5}/10\right)\,w_k\,.  \label{iii20}
\end{equation}
Then,
\begin{equation}
\mathcal{W}[v_{i,k},v_k]= \exp\left(-\beta_k\,z^{-5}/5\right)\,\mathcal{W}[w_{i},w_k]\,.  \label{iii21}
\end{equation}
For the left hand side we have the doubly infinite series
\begin{equation}
\mathcal{W}[v_{i,k},v_k]\sim \sum_{n=-\infty}^\infty \gamma_n^{(i,k)}\,z^{n+\nu_i+\rho_k}\,,  \label{iii22}
\end{equation}
with coefficients
\begin{equation}
\gamma_n^{(i,k)}=\sum_{m=0}^\infty b_{m,k}\big(-\beta_k\,c_{n-m+6,i}+(-n+2m-1-\nu_i+\rho_k)\,c_{n-m+1,i}\big),  \label{iii23}
\end{equation}
which obey the fifth order difference equation
\begin{equation}
(n-5+\nu_i+\rho_k)\,\gamma_{n-5}^{(i,k)}-\beta_k\,\gamma_n^{(i,k)} = 0\,.  \label{iii24}
\end{equation}
Five independent solutions of this difference equation are constituted by the coefficients of the five Heaviside's exponential series
\begin{equation}
\exp\left(-\beta_k\,z^{-5}/5\right) \sim \sum_{n=-\infty}^{\infty}\frac{\left(-\beta_k\,z^{-5}/5\right)^{n+\delta_{L}^{(i,k)}}}
{\Gamma(n+1+\delta_{L}^{(i,k)})}, \qquad  L=0,1,\ldots, 4\,, \label{iii25}
\end{equation}
with
\begin{equation}
\delta_L^{(i,k)}=(-\nu_i-\rho_k+L)/5\,.  \label{iii26}
\end{equation}
Then, analogously to Eqs. (\ref{iii16}) and (\ref{iii19}), one has
\begin{eqnarray}
\mathcal{W}[w_{i},w_5]&=&\sum_{L=0}^4\frac{\Gamma (n+1+\delta_L^{(i,5)})}{(-\beta_5/5)^{n+\delta_L^{(i,5)}}}\,\gamma_{-5n-L}^{(i,5)}\,, \label{iii27} \\
\mathcal{W}[w_{i},w_6]&=&(-1)^n\sum_{L=0}^4 \,\cos (\delta_L^{(i,6)}\pi)\, \frac{\Gamma (n+1+\delta_L^{(i,6)})}{(\beta_6/5)^{n+\delta_L^{(i,6)}}}\,\gamma_{-5n-L}^{(i,6)}\,. \label{iii28}
\end{eqnarray}
Now it is immediate to calculate the connection factors $T_{i,5}$ and $T_{i,6}$ by means of Eq. (\ref{iii3}).

\section{Bound states}

By using the above described procedure, we have determined the values of $\varepsilon$ which are solution of Eq. (\ref{ii20}) for different intensities of the potential in the range $0<\lambda<100$ and for five values of the angular momentum, $l=0, 1, \ldots, 4$. The results are shown graphically in Figure 1.
\begin{figure}
\includegraphics{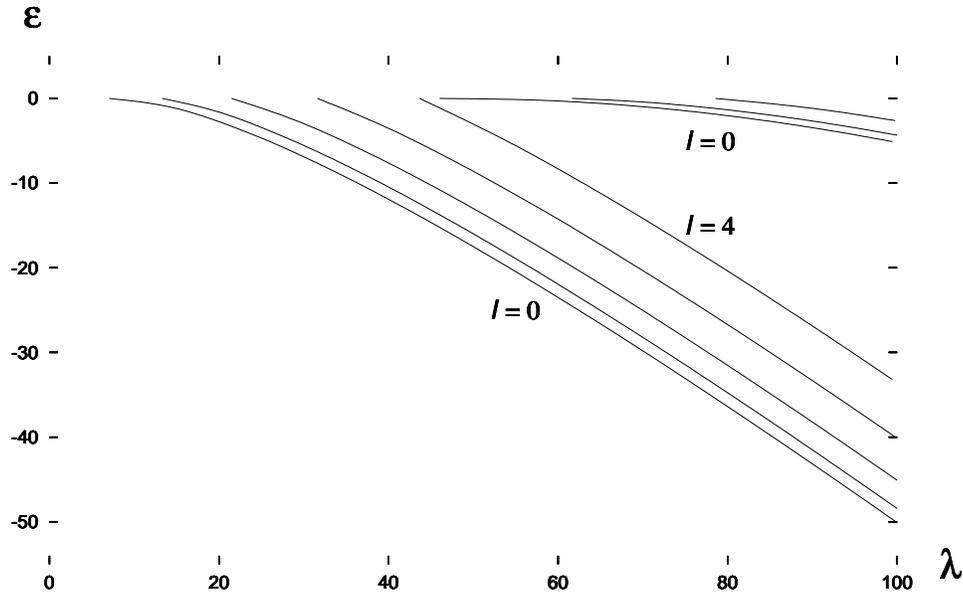}
\caption{Energies of bound states in the Lennard-Jones potential. The graph shows the energies of the lowest states of angular momentum $l=0, 1, 2, 3, 4$, and the first excited states with $l=0, 1, 2$, for a varying intensity of the potential in the range $0\leq\lambda\leq 100$. The curves corrresponding to the lowest states with $l=5, 6, 7$, intersect those shown of the first excited states and have been omitted for the sake of clarity of the figure.}
\label{fig:1}
\end{figure}

Besides the energies of the bound states, our procedure gives also their wave functions. For the values of $\varepsilon$ satisfying Eq. (\ref{ii20}), $A_1$ and $A_2$ can be determined, save for a common arbitrary multiplicative constant, by using any one of Eqs. (\ref{ii19}). To fix the arbitrary constant, we may impose, for instance, that
\begin{equation}
A_1\,T_{1,3}+A_2\,T_{2,3}=1.   \label{iv1}
\end{equation}
Then
\begin{equation}
A_1=\frac{T_{2,4}}{T_{1,3}\,T_{2,4}-T_{2,3}\,T_{1,4}}\,, \qquad  A_2=\frac{-\,T_{1,4}}{T_{1,3}\,T_{2,4}-T_{2,3}\,T_{1,4}}\,, \label{iv2}
\end{equation}
and, in view of Eqs. (\ref{ii16}) and (\ref{ii5}), the wave function of the bound state becomes
\begin{equation}
w_{\rm phys}(z)=\mathcal{N}\left(A_1\,z^{\nu_1}\sum_{n=-\infty}^{\infty} c_{n,1}\,z^n+A_2\,z^{\nu_2}\sum_{n=-\infty}^{\infty} c_{n,2}\,z^n\right),  \label{iv3}
\end{equation}
$\mathcal{N}$ being a normalization constant such that
\begin{equation}
\int_0^\infty dz\,|w_{\rm phys}(z)|^2=r_{\mbox{e}}^{-1}\,.   \label{iv4}
\end{equation}
For large values of $z$, the series in Eq. (\ref{iv3}) converge slowly and are not convenient for the computation of $w_{\rm phys}(z)$. In this case, it is preferable to use the asymptotic expansion
\begin{equation}
w_{\rm phys}(z)\sim\mathcal{N}\,\exp\left(\alpha_{3}\,z\right) \,\sum_{m=0}^\infty a_{m,3}\,z^{-m}\,, \qquad z\to\infty\,, \label{iv5}
\end{equation}
stemming from
\begin{equation}
w_{\rm phys}(z) \sim \mathcal{N}\left((A_1\,T_{1,3}+A_2\,T_{2,3})\,w_3(z)+(A_1\,T_{1,4}+A_2\,T_{2,4})\,w_4(z)\right),
\end{equation}
bearing in mind Eqs. (\ref{ii19}) and (\ref{iv1}) and the expansion in Eq. (\ref{ii7}). For the same reason, one should use the asymptotic expansion
\begin{equation}
w_{\rm phys}(z)\sim\mathcal{N}\,\left(A_1\,T_{1,5}+A_2\,T_{2,5}\right)\,\exp\left(\beta_{5}\,z^{-5}/5\right) \,\sum_{m=0}^\infty b_{m,5}\,z^{m}\,, \qquad z\to 0\,, \label{iv6}
\end{equation}
in the neighbourhood of the origin.

We have obtained, by way of illustration, the parameters of the four existing bound states in a potential of intensity $\lambda=40$. Tables 1 to 4 show the values of the energy, the indices $\nu_i$ of the Floquet solutions, the connection factors, and the coefficients $A_i$ to be substituted in Eq. (\ref{iv3}), for each one of those bound states.
For the determination of the indices $\nu_i$ and the coefficients $c_{n,i}$ of the Floquet solutions, we used the Newton iteration method, to be recalled in the Appendix. We benefited from the subroutines \verb"bandec" and \verb"banbks" \cite[pp. 45--46]{pres} to obtain the initial values, and from \verb"ludcmp" and \verb"lubksb" \cite[pp. 38--39]{pres} in the iteration process. Double precision \textsc{Fortran} was used in the computation. The iteration was stopped when the correction in the absolute value of $\nu_i$ became less than $10^{-13}$. Usually, two or three iterations were enough. Simultaneously, the coefficients $c_{n,i}$, with $-360\leq n\leq 360$, were obtained. (Due to the fact that Eq. (\ref{ii6}) relates coefficients with subindexes of the same parity, the ambiguity in the definition of $\nu_i$, mentioned in Subsection 2.1, allows one to cancel all coefficients $c_{n,i}$ with odd $n$.)
According to the condition (\ref{extra2}), to be justified in the Appendix, the indices of the Floquet solutions either are real or, being complex, have opposite imaginary parts. In this case, thanks to the ambiguity in the definition of the $\nu_i$, one may choose them to be complex conjugate to each other. Then, $w_2$, $T_{2,j}$, $T_{2,k}$ and $A_2$ are the complex conjugate of, respectively,  $w_1$, $T_{1,j}$, $T_{1,k}$ and $A_1$. Consequently, $w_{\rm phys}(z)$ becomes real.

A word of caution about the computation of the wave function is in order. Our double precision calculations have revealed that Eq. (\ref{iv3}), with the series truncated to $\sum_{n=-200}^{200}$, allows one to obtain values of $w_{\rm phys}(z)$ with eight correct significant digits whenever roughly $0.7< z < 3.0$, whereas the asymptotic expansions in Eqs. (\ref{iv5}) and (\ref{iv6}) become useful for $z>4.5$ and $z< 0.4$, respectively. Therefore, double precision is not sufficient for a computation of the values of $w_{\rm phys}(z)$ in the whole interval $0<z<\infty$. Quadruple precision calculations, instead, provide with satisfactory results.

\begin{table}
\caption{Parameters of the ground state in a Lennard-Jones potential of intensity $\lambda=40$.}
\label{tab:1}
\begin{tabular}{cc}
\hline\noalign{\smallskip}
angular momentum & $l=0$ \\
energy & $\varepsilon = -\,11.909183$ \\
$\nu_1$ &  $0.5\,-\,3.31231657\,i$ \\
$T_{1,3}$ & $-\,0.10275762$E$+03\,-\,0.20083284$E$+03\;i$  \\
$T_{1,4}$ & $-\,0.12151177$E$-01\,+\,0.62172400$E$-02\;i$ \\
$T_{1,5}$ & $-\,0.13871649$E$+04\,-\,0.26958725$E$+04\;i$  \\
$T_{1,6}$ & $\ \ \,0.49335027$E$-03\,-\,0.25242634$E$-03\;i$  \\
$A_1$ & $-\,0.10095465$E$-02\,+\,0.19730906$E$-02\;i$  \\
\noalign{\smallskip}\hline
\end{tabular}
\end{table}
\begin{table}
\caption{Parameters of the first excited state in a Lennard-Jones potential of intensity $\lambda=40$.}
\label{tab:2}
\begin{tabular}{cc}
\hline\noalign{\smallskip}
angular momentum & $l=1$ \\
energy & $\varepsilon = -\,10.465279$ \\
$\nu_1$ &  $0.5\,-\,2.99607877\,i$ \\
$T_{1,3}$ & $-\,0.56554657$E$+02\,-\,0.13235260$E$+03\;i$  \\
$T_{1,4}$ & $-\,0.21626362$E$-01\,+\,0.92410081$E$-02\;i$ \\
$T_{1,5}$ & $-\,0.11876972$E$+04\,-\,0.27655908$E$+04\;i$  \\
$T_{1,6}$ & $\ \ \,0.52897583$E$-03\,-\,0.22603293$E$-03\;i$  \\
$A_1$ & $-\,0.13650231$E$-02\,+\,0.31945090$E$-02\;i$   \\
\noalign{\smallskip}\hline
\end{tabular}
\end{table}
\begin{table}
\caption{Parameters of the second excited state in a Lennard-Jones potential of intensity $\lambda=40$.}
\label{tab:3}
\begin{tabular}{cc}
\hline\noalign{\smallskip}
angular momentum & $l=2$ \\
energy & $\varepsilon = -\,7.629685$ \\
$\nu_1$ &  $0.5\,-\,2.26050463\,i$ \\
$T_{1,3}$ & $-\,0.75165112$E$+01\,-\,0.49967087$E$+02\;i$  \\
$T_{1,4}$ & $-\,0.82325669$E$-01\,+\,0.12384188$E$-01\;i$ \\
$T_{1,5}$ & $-\,0.40737562$E$+03\,-\,0.28987415$E$+04\;i$  \\
$T_{1,6}$ & $\ \ \,0.62067701$E$-03\,-\,0.93367973$E$-04\;i$  \\
$A_1$ & $-\,0.14719741$E$-02\,+\,0.97851589$E$-02\;i$   \\
\noalign{\smallskip}\hline
\end{tabular}
\end{table}
\begin{table}
\caption{Parameters of the third excited state in a Lennard-Jones potential of intensity $\lambda=40$.}
\label{tab:4}
\begin{tabular}{cc}
\hline\noalign{\smallskip}
angular momentum & $l=3$ \\
energy & $\varepsilon = -\,3.530328$ \\
$\nu_1$ &  $0.5\,-\,0.59466296\,i$ \\
$T_{1,3}$ & $\ \ \,0.42589066$E$+01\,-\,0.28071593$E$+01\;i$  \\
$T_{1,4}$ & $-\,0.88803914$E$+00\,-\,0.13472965$E$+01\;i$ \\
$T_{1,5}$ & $\ \ \,0.35206173$E$+04\,+\,0.56914086$E$+03\;i$  \\
$T_{1,6}$ & $\ \ \, 0.51239569$E$-03\,+\,0.77738566$E$-03\;i$  \\
$A_1$ & $\ \ \,0.81844039$E$-01\,+\,0.53945596$E$-01\;i$   \\
\noalign{\smallskip}\hline
\end{tabular}
\end{table}

\section{Critical values of the intensity}

It may be interesting to know the values of $\lambda$ for which a new bound state (of zero energy) appears. Our method of solution of the Schr\"odinger equation is also applicable in this case, but in a much simpler form. For zero energy, the singular point at infinity is a regular one and the basic Floquet solutions of the general case are replaced by Frobenius solutions whose coefficients can be obtained trivially. The procedure in this case is the same used to obtain the scattering length \cite{gom3}. In fact, as it is well known, the presence of a new bound state of zero energy is revealed by a pole in the scattering length. We report, in Table 5, some critical values of the intensity $\lambda$ for different values of the angular momentum $l$.
\begin{table}
\caption{Lowest values of the of intensity $\lambda$ of the Lennard-Jones potential Eq. (\ref{i1}) for which a new bound state of angular momentum $l$ appears.}
\label{tab:5}
\begin{tabular}{rrrrrr}
\hline\noalign{\smallskip}
$l=0$ & $l=1$ & $l=2$ & $l=3$ & $l=4$ & $l=5$  \\
\noalign{\smallskip}\hline\noalign{\smallskip}
7.04314 & 13.29573 & 21.48500 & 31.60949 & 43.66864 & 57.66218  \\
46.61703 & 61.64985 & 78.58395 & 97.43067 & 118.19665 & 140.88604  \\
121.28583 & 145.10984 & 170.82095 & 198.43005 & 227.94507 & 259.37186  \\
231.08863 & 263.70031 & 298.19340 & 334.57660 & 372.85712 & 413.04084  \\
376.02780 & 417.42555 & 460.70191 & 505.86378 & 552.91734 & 601.86799  \\
\noalign{\smallskip}\hline
\end{tabular}
\end{table}

\section{Final comments}

We have shown the applicability of our method for obtaining global solutions of the Schr\"odinger equation in the case of bound states in a (12,6) Lennard-Jones potential. The method can be similarly applied to any other Lennard-Jones-type potential, whatever exponents in the attractive and repulsive terms. The physical solution results as a determined linear combination of the two Floquet solutions and its asymptotic expansion at the singular points is proportional to the respective regular Thom\'e solutions.

Given a value of the intensity $\lambda$ of the potential, a study of the indices $\nu_i$ of the Floquet solutions reveals that they are real for small energy. They may be taken in the interval $0\leq\nu_i\leq 1$, with $\nu_2=1-\nu_1$. As the energy increases, $\nu_1$ increases and $\nu_2$ decreases, both approaching the value 1/2 for a certain energy. As $\nu_1=\nu_2=1/2$, only one multiplicative solution exists: any other independent solution of the Schr\"odinger equation contains logarithmic terms. Increasing the energy makes both $\nu_1$ and $\nu_2$ to become complex, with fixed common real part equal to $1/2$ and opposite imaginary parts increasing with the energy. The physical wave function, however, may be taken real by adjusting the arbitrary global phase.

Special mention deserve the critical values of the intensity discussed in Section 5. Our Table 5 allows one to know immediately the number of states, of each angular momentum, bounded by a potential of given intensity.

\section*{Appendix}

We have mentioned in Subsection 2.1 that the computation of the indices and coefficients of the Floquet solutions can be treated as a nonlinear eigenvalue problem, whose solution we are going to consider in this Appendix. Along it we will omit, for brevity, the subindex $i$ in $\nu_i$ and $c_{n,i}$. The condition in Eq. (\ref{ii5}) implies that
\begin{equation}
\lim_{n\to\pm\infty}|c_n|=0,  \label{auno}
\end{equation}
which allows one to truncate the infinite set of equations (\ref{ii6}) and to restrict the label
$n$ to the interval $-M \leq n \leq N$, both $M$ and $N$ being positive integers large enough to guarantee
that the solution of the truncated problem does not deviate significantly from that of the original infinite one. Algorithms to solve finite-order problems have been discussed by Ruhe \cite{ruhe}. Here we recall the Newton iteration method suggested by Naundorf \cite{naun}. The procedure consists in moving from an approximate solution,
$\{\nu^{(i)}, c_n^{(i)}\}$, to another one,  $\{\nu^{(i+1)}, c_n^{(i+1)}\}$, by solving the system of equations
\begin{eqnarray}
& \hspace{-30pt}\varepsilon\,c_{n-2}^{(i+1)} + \left[\big(n\! +\! \nu^{(i)}\big)\big(n\! -\! 1\! +\! \nu^{(i)}\big)-l(l+1)\right]c_n^{(i+1)} + 2\lambda\,c_{n+4}^{(i+1)} - \lambda\,c_{n+10}^{(i+1)}
\nonumber \\
& +\,\big(2n\! -\! 1\! +\! 2\nu^{(i)}\big)c_n^{(i)}\big(\nu^{(i+1)} - \nu^{(i)}\big)  =  0, \qquad n=-{M}, \ldots, -1, 0, 1, \ldots, {N},
\label{ados} \\
&\hspace{-132pt}\sum_{n=-{M}}^{N}{c_n^{(i)}}^*c_n^{(i+1)} =  1,  \label{atres}
\end{eqnarray}
that results, by linearization \cite{naun}, from (\ref{ii6}) and from the truncated normalization condition
\[
\sum_{n=-M}^{N}\left| c_{n}\right|^2 = 1.
\]
Obviously, the values of $c_m^{(i)}$ with $m<-{M}$ or $m>{N}$ entering in some of Eqs. (\ref{ados}) should be taken
equal to zero, in accordance with the truncation done.  The iteration process is stopped when the
difference between consecutive solutions, $\{\nu^{(i)},\,c_n^{(i)}\}$ and $\{\nu^{(i+1)},\,c_n^{(i+1)}\}$ is satisfactory. The resulting values of $\nu$ and $c_n$ may serve as initial values for a new iteration process, with
larger values of ${M}$ and ${N}$, to check the stability of the solution.

Of course, the Newton method just described needs initial values $\{\nu^{(0)}, c_n^{(0)}\}$ not far from the true solution.
The two different values of $\nu$ can be obtained from the two eigenvalues
\begin{equation}
 \exp (2i\pi\nu_i)\,,    \qquad i=1, 2\,,   \label{acuatro}
\end{equation}
of the circuit matrix ${\bf C}$ \cite{waso} for the singular point at $z=0$.
The entries of that matrix can be computed by numerically integrating Eq. (\ref{ii4}) on the unit circle, from $z=\exp (0)$ to $z=\exp (2i\pi)$, for two independent sets of initial values. If we consider two solutions, $w_a(z)$ and $w_b(z)$, obeying, for instance, the conditions
\begin{eqnarray}
w_a(\mbox{e}^0)=1,\qquad w_a^\prime(\mbox{e}^0)=0, \nonumber \\
w_b(\mbox{e}^0)=0,\qquad w_b^\prime(\mbox{e}^0)=1, \nonumber
\end{eqnarray}
then
\begin{eqnarray}
C_{11}=w_a(\mbox{e}^{2i\pi}), \qquad C_{12}=w_b(\mbox{e}^{2i\pi}), \nonumber \\
C_{21}=w_a^\prime(\mbox{e}^{2i\pi}), \qquad C_{22}=w_b^\prime(\mbox{e}^{2i\pi}), \nonumber
\end{eqnarray}
and
\begin{equation}
\nu=\frac{1}{2i\pi}\,\ln \left[\frac{1}{2}\left( C_{11}+C_{22}\pm\sqrt{\left(C_{11}\! -\! C_{22}\right)^2+4C_{12}C_{21}}\right)\right].
\label{acinco}
\end{equation}
The two signs in front of the square root produce two different values for $\nu$, unless the parameters $\lambda$ and $\varepsilon$
in Eq. (\ref{ii4}) be such that $\left(C_{11}-C_{22}\right)^2+4C_{12}C_{21}=0$, in which case only one
multiplicative solution appears, any other independent solution containing logarithmic terms.
The ambiguity in the real part of $\nu$ due to the multivaluedness of the logarithm in the right
hand side of (\ref{acinco}) reflects the fact already mentioned that the indices $\nu$ are not uniquely defined.
Notice that
\begin{equation}
\exp (2i\pi\nu_1)\, \exp (2i\pi\nu_2)=\det {\bf C}=\mathcal{W}[w_a,w_b]=1   \label{extra1}
\end{equation}
and, therefore,
\begin{equation}
\nu_1+\nu_2=0\quad (\mbox{mod} \;1).  \label{extra2}
\end{equation}
This may serve as a test for the integration of Eq. (\ref{ii4}) on the unit circle.

Although Eq. (\ref{acinco}) is exact, the $C_{mn}$ are obtained by numerical integration of a differential equation  and are not sufficiently precise. The resulting values of $\nu$ may only be considered as starting values, $\nu^{(0)}$, for the Newton iteration process.
As starting coefficients $c_n^{(0)}$ one may use the solutions of the homogeneous system
\begin{eqnarray}
&\varepsilon\,c_{n-2}^{(0)}+
\left[(n\! +\! \nu^{(0)})(n\! -\! 1\! +\! \nu^{(0)})-l(l+1)\right]c_{n}^{(0)} +2\lambda\,c_{n+4}^{(0)}-\lambda\,c_{n+10}^{(0)}=0\,, \nonumber \\
&n= -M,  \ldots, -1, 0, 1, \ldots, N,   \label{aseis}
\end{eqnarray}
with the already mentioned truncated normalization condition
\begin{equation}
\sum_{n=-M}^N |c_{n}^{(0)}|^2 = 1.  \label{asiete}
\end{equation}

\section*{Acknowledgments}
Financial support from Departamento de Ciencia, Tecnolog\'{\i}a y Universidad del Gobierno de Arag\'on (Project E24/1) and Ministerio de Ciencia e Innovaci\'on (Project MTM2009-11154) is gratefully acknowledged.

\end{document}